\documentclass[11pt]{article}
\usepackage{pdfsync}
\usepackage{graphicx} 
\usepackage{subfig}
\usepackage{color}
\usepackage{amsfonts}
\usepackage{amssymb}
\usepackage{epsfig}
\usepackage{dcolumn}
\usepackage{booktabs}
\usepackage{multirow}
\usepackage{makecell}
\usepackage{boldline}
\usepackage{marvosym}

\usepackage[colorlinks]{hyperref}
\hypersetup{
    colorlinks=true,
    linkcolor=blue,
    filecolor=magenta,  
    urlcolor=[rgb]{0.00,0.00,0.50},    
    citecolor=red,
    }

\usepackage{url}

\advance \textheight by \topskip
\usepackage{amsmath}
\usepackage[english]{babel}
\makeatletter
 \@addtoreset{equation}{section}
\makeatother

\usepackage{amsmath,amssymb}
\makeatletter \@addtoreset{equation}{section}
\makeatother
\def\be{\begin{equation}}

\def\be{\begin{equation}}
\def\ee{\end{equation}}

\def\R{\mathbb R}

\def\C{\mathbb C}

\usepackage{amsmath,amssymb}
\def\bea{\begin{eqnarray}}
\def\eea{\end{eqnarray}}
\def\barray{\begin{array}}
\def\earray{\end{array}}

\begin{document}

\title{
{\bf 
Schwarzian derivative treatment of  the quantum second-order supersymmetry anomaly,
and coupling-constant metamorphosis
}}

\author{{\bf Mikhail S. Plyushchay}  \\
[8pt]
{\small \textit{
Departamento de F\'{\i}sica,
Universidad de Santiago de Chile, Casilla 307, Santiago 2,
Chile  }}\\
[4pt]
 \sl{\small{E-mail:  
\textcolor{blue}{mikhail.plyushchay@usach.cl}
}}
}
\date{}
\maketitle

\begin{abstract}
A  canonical quantization scheme applied to a
classical supersymmetric system with  quadratic in momentum 
supercharges gives rise to a quantum anomaly  problem described by a 
specific   term  to be quadratic in Planck constant.
We reveal  a close  relationship  between  the anomaly 
and the Schwarzian derivative, and  specify
a quantization prescription which generates  the anomaly-free  supersymmetric 
quantum system with second order supercharges.  We also  discuss the phenomenon of
 a coupling-constant metarmorphosis that  associates quantum 
systems with the first-order supersymmetry to the systems with the second-order supercharges.
\end{abstract}

\vskip.5cm\noindent

\section{Introduction}
A supersymmetric quantum mechanical system
is characterized by supercharges 
which can be differential operators of the first \cite{Witten,Cooper,Junker,CKSb} or higher 
 \cite{AnIoSp,AnCaDeIo,Fernan,Bagchi,Plyu1,KliPly,IofNis,CJNP,CJPAdS,BerUss1,BerUss2}
 order. 
At the classical level, the corresponding 
supercharges are linear or nonlinear functions 
of the momentum \cite{Plyu1,KliPly}. 
In a simplest case of one dimension
and non-extended supersymmetry, 
the  Hamiltonian and two supercharges of a system
are given in terms of 
one function which is usually referred to as a 
superpotential. 
The higher order supercharges generate non-standard
supersymmetry algebras at the classical and quantum 
levels.  The algebraic
structure of such a higher order 
supersymmetry, see Eqs. 
(\ref{QQHclas}) and 
(\ref{QQquant}) below for the case of the second-order supersymmetry,
 resembles 
a structure of the finite W-algebras
\cite{Rocek,Walgebras}
in which classical Poisson bracket or 
commutator of generating elements 
is proportional to a  polynomial
in them.
In the case of finite W-algebras\footnote{The 
interest to finite W-algebras was 
generated by investigations of analogous 
infinite-dimensional algebraic structures
appearing in conformal field theory and 
integrable systems \cite{DrinSok,Drinfeld, Jimbo,Zamolod,FatLuk,BouSch,Dickey}.}
the polynomial in momentum integrals of motion
are related with and reflect effectively peculiar properties of a 
system, for instance,  
they explain an additional (`accidental') degeneracy in 
the quantum spectrum and a closed character of  
classical trajectories 
in multi-dimensional oscillator systems  with 
commensurable  frequencies  and in Kepler (hydrogen model) 
problem \cite{Walgebras}. 
Similarly to the systems of bosonic oscillators with 
commensurable  frequencies, nonlinear supersymmetry 
appears in superconformal mechanics at the special 
values of the boson-fermion coupling constant \cite{PlyLei}.
The  higher-order supercharges also emerge 
in quantum mechanical systems with soliton potentials 
and in finite-gap systems intimately
related to nonlinear completely integrable systems
 \cite{CJNP,CJPAdS,CorNiePly,CorrDunPl,CorrLechPly}. 
They generate there exotic supersymmetric 
structure associated with factorization of 
the higher-derivative 
Lax-Novikov integrals of motion which 
underlie reflectionless  and finite-gap properties
of the corresponding quantum mechanical 
systems.  Such higher-order 
exotic supersymmetric structure was employed 
recently in \cite{ChirAsym} for the construction of 
new solutions for the Korteweg-de Vries and 
modified Korteweg-de Vries equations in the form of  
different types of soliton defects 
in crystalline backgrounds which reveal a peculiar 
dynamics with the chiral asymmetry.
The second-order supersymmetry based on the 
confluent Darboux-Crum transformations 
was used in \cite{CorJacPly} for 
the design of the PT-symmetric nonlinear optical  systems with 
completely invisible periodicity defects.
Another interesting  application 
of nonlinear supersymmetry is related with
a new, recently discovered class of 
exceptional orthogonal polynomials 
describing  quantum mechanical systems  with
rational extensions of harmonic, isotonic and  
Darboux-P\"oschl-Teller potentials 
\cite{Odake,Quesne11,DavGom}.


In spite of a similarity, there is
an  essential difference between the cases 
of supersymmetries with linear and higher-order
supercharges.
Classically, supersymmetries of these two types
are realized on the basis of a superpotential which can be 
an arbitrary function. Under attempt 
of quantization in the case of
nonlinear supersymmetry,  there appears a  problem
of the quantum anomaly: a usual canonical quantization 
produces a quantum system with the broken 
supersymmetry in which quantum analogs of supercharges are not
anymore integrals of motion.

In this work  we investigate 
the problem of the quantum anomaly in a simplest
nontrivial case of supersymmetry with quadratic in momentum 
supercharges. 
We find out that in this case there exists a  universal `remedy' 
against the quantum anomaly `disease' 
provided by the Schwarzian derivative.
Bearing in mind that the Schwarzian derivative plays also a key
role in a coupling-constant metamorphosis
and duality between quantum systems \cite{Hieat,HietGram,CabGal,Cariglia},
we study this phenomenon in the
context of supersymmetry to show that it 
provides here a specific duality between the systems with the 
second and first order supercharges.  
 
The paper is organized as follows. 
In the next Section \ref{AnomalySec} we first reproduce a general structure
of the classical and quantum supersymmetric systems 
with the second order supercharges. We analyze then different 
 representations of the quantum superpartner Hamiltonians 
and supercharges and identify a specific term to be quadratic in Planck
constant which is responsible for the quantum anomaly.
We reveal its close relationship with the Schwarzian derivative, 
and, finally, specify the anomaly-free quantization prescription
for classical systems with  the second order supersymmetry.
Section \ref{Metamor} is devoted to investigation 
of the coupling-constant metamorphosis and associated 
duality phenomenon in the context of the second order supersymmetry.
In Section \ref{SchwSchr} we consider the form-invariant transformations
in a stationary Schr\"odinger equation to understand better 
 the roots of the Schwarzian derivative 
`remedy' for the quantum anomaly `disease' 
as well as to see a  hidden role played by the Schwarzian  
in  the coupling-constant  metamorphosis mechanism.  
In Section \ref{Examples} the obtained general results are illustrated 
by some simple examples. 
The last Section \ref{Conclusion} is devoted to concluding remarks and outlook.
In Appendix,  we describe a most general structure of the 
factorizing operators which appear in quantum systems with 
the second order supersymmetry.

 
\section{Quantum anomaly and Schwarzian derivative treatment }\label{AnomalySec}

Consider a classical particle system described by the Hamiltonian
\be\label{mathH}
\mathcal{H}=\frac{1}{2}\left(p^2+v(x)\right) +\nu(x)N\,.
\ee
Here 
$N=\theta^+\theta^-$,
$\theta^+$ and $\theta^-=(\theta^+)^*$ are 
complex Grassmann variables,  $v(x)$ and $\nu(x)$
are real functions to be fixed below, and we use the units 
with mass $m=1$.
Nontrivial  
Poisson brackets $\{x,p\}=1$, 
$\{\theta^+,\theta^-\}=-i$  give rise to
the equations of motion $\dot{x}=p$, $\dot{p}=-\frac{1}{2}v'-\nu'N$, 
$\dot{\theta}^\pm=\pm i\nu\theta^\pm$, and we find that  
system (\ref{mathH})
has an obvious even nilpotent   integral of motion $N$.
Let us require that, in addition,  there exists a pair of the odd integrals of motion 
quadratic in $p$, 
\be
	\mathcal{Q}_+=\frac{1}{2}\left(
	-p^2+2iW(x)p+g(x)\right)\theta^+\,,
	\qquad
	\mathcal{Q}_-=(\mathcal{Q}_+)^*\,,
\ee
where $W(x)$ and $g(x)$ are some functions.
The requirement  $\dot{\mathcal{Q}}_+=0$
yields   three equations
which correspond to  the nulling conditions for
 coefficients at $p^2$, $p$ and $p^0$\,:
\be\label{3eq}
	-\nu +2W'=0\,,\qquad
	 \frac{1}{2}v'+\frac{1}{2}g'-\nu W=0\,,\qquad 
	\nu g-Wv'=0\,.
\ee
{}From the first two equations we have
$\nu=2W'$ and $v=2W^2-g+a$,
where $a$ is a real integration constant.
Using these relations in the third equation from (\ref{3eq})
and multiplying it 
by $W$, we obtain the equivalent equation
$\frac{d}{dx}\left[W^2(g-W^2)\right]=0$.
{}From here,  $g(x)$ is fixed in terms of $W(x)$, 
$g(x)=W^2(x)+\frac{C}{W^2(x)}$,
where $C$ is an integration constant.
Introducing the notation
$z=W(x)+ip,$
we conclude that 
the system described by the Hamiltonian 
\be\label{mathHcl}
	\mathcal{H}=\frac{1}{2}\left(z z^* -\frac{C}{W^2}+4W'N \right)+a\,,
\ee
possesses the two odd integrals of motion 
\be\label{Q+clas}
	\mathcal{Q}_+=\frac{1}{2}\left(z^2+\frac{C}{W^2}\right)\theta^+\,\qquad
	\text{and}\qquad  \mathcal{Q}_-=(\mathcal{Q}_+)^*\,,
\ee
which are the quadratic in momentum classical  supercharges.
Function $W(x)$ and constant $C$ 
ought  to be real  in order Hamiltonian (\ref{mathHcl}) 
be real. We shall refer to the function $W(x)$ as a superpotential.
The integrals $\mathcal{H}$, $\mathcal{Q}_+$, 
$\mathcal{Q}_-$ and  $N$ generate
nonlinear  (quadratic) classical superalgebra with   the only nontrivial 
Poisson bracket relations  $\{N,\mathcal{Q}_\pm\}=\mp i 
\mathcal{Q}_\pm$ and 
\be\label{QQHclas}
\{\mathcal{Q}_+,\mathcal{Q}_-\}=-i\left((\mathcal{H}-a)^2 +C\right)\,.
\ee
The constant $C$ plays a role of 
the central charge (alongside with $\mathcal{H}-a$) of this 
superalgebra.
The transformation 
\be\label{classymd}
	W\rightarrow -W\,,\qquad 
	 \theta^+\rightarrow \theta^-\,,\qquad
	\theta^-\rightarrow \theta^+
\ee  
corresponds here to a discrete symmetry of the system
under which $\mathcal{H}\rightarrow \mathcal{H}$,
$\mathcal{Q}_+\leftrightarrow 
\mathcal{Q}_-$, $N\rightarrow -N$.
\vskip0.1cm


Let us quantize  the system.
Taking $\hat{\theta}_\pm={\sqrt{\hbar}}\,\sigma_\pm$,
where 
$\sigma_\pm=\frac{1}{2}(\sigma_1\pm i\sigma_2)$
are linear combinations  of the Pauli matrices,
we have $[\hat{\theta}_+,\hat{\theta}_-]_+=\hbar$
in correspondence with the Grassmann nature of 
the variables $\theta^\pm$ 
and their classical Poisson brackets.
As a quantum analog of $N$ we take
$\hat{N}=\frac{1}{2}(\hat{\theta}_+\hat{\theta}_- - 
\hat{\theta}_-\hat{\theta}_+)=
\frac{1}{2}\hbar\sigma_3$. 
To construct quantum analogs of $\mathcal{H}$ and $\mathcal{Q}_+$ we proceed from 
the classical relations $2(\mathcal{H}-a)=p^2+W^2-\frac{C}{W^2}+4W'N$,
$2\mathcal{Q}_+=\left((W+ip)^2+\frac{C}{W^2}\right)\theta_+$,
and substitute in them directly  $\theta_\pm$ and $N$
 for their quantum analogs, and $p$ for $-i\hbar \frac{d}{dx}$.
 The operators obtained by application of 
 this direct  quantization prescription 
 we denote by  $\hat{\mathcal{H}}_{d}$
 and $\hat{\mathcal{Q}}_{+d}$, and
find that  $[\hat{\mathcal{H}}_d,\hat{\mathcal{Q}}_{+d}]=0$ 
if and only if  $W=(a_1x+a_0)^2$,
where $a_1$ and $a_0$ are some constants.
In the case of the superpotential of 
any other form, this commutator is nonzero, and we face the problem of
the quantum anomaly \cite{Plyu1,KliPly},  see below. It is known, however, 
that supersymmetry with the second order supercharges
can be realized at the quantum level 
for superpotential $W(x)$  to be a function 
of an arbitrary form. 
We arrive then at a  rather natural question whether  some another  
quantization scheme for  the classical system 
(\ref{mathHcl}),  (\ref{Q+clas}) exists which 
would produce an anomaly-free 
quantum system with the second order supercharges
given in terms of an arbitrary function $W(x)$.
To answer the question we shall construct,  
analogously to the  classical case presented above, 
a quantum system described by   the matrix
Hamiltonian operator $\hat{\mathcal{H}}$
by demanding  the existence of the  second order supercharge
operators $\hat{\mathcal{Q}}_\pm$. Then we
present the obtained $\hat{\mathcal{H}}$ and $\hat{\mathcal{Q}}_\pm$ 
 in terms of  the  factorizing first order differential
operators, and compare the structure  of
$\hat{\mathcal{H}}$ and $\hat{\mathcal{Q}}_\pm$ 
with that of  $\hat{\mathcal{H}}_d$ and 
$\hat{\mathcal{Q}}_{\pm d}$. This  will
allow us to identify  the anomaly-free prescription
for quantization of the classical system.

Before we proceed further, let us note  
that if we  would choose another 
simple ordering prescription 
under construction of the
quantum analog of (\ref{mathHcl}) by substituting in it
the classical term $zz^*$, for instance, 
 for $(W+\hbar \frac{d}{dx})(W-\hbar\frac{d}{dx})$,
this would  produce additional term $\hbar W'$. 
This  change can be compensated by choosing 
another ordering 
prescription in the construction of the quantum analog of $N$.
Indeed, classically we can write $N=\sin^2\beta \, \theta^+\theta^-
-\cos^2\beta\, \theta^-\theta^+$ with arbitrary real parameter $\beta$. 
Based on this expression, we could take as a quantum analog
$\hat{N}=\sin^2\beta\,\hat{\theta}^+ \hat{\theta}^- -
\cos^2\beta\, \hat{\theta}^- \hat{\theta}^+$.
The choice $\beta=\frac{\pi}{4}$ corresponds to the antisymmetric 
ordering we chose  above with  
$\hat{N}=\frac{1}{2}\hbar\sigma_3$. In the present case, the choice 
of the quantum ordering with $\beta=\frac{\pi}{6}$ 
yields as a result  the same 
quantum Hamiltonian   $\hat{\mathcal{H}}_{d}$. 

As we shall see, simple changes in the 
ordering of the non-commuting factors under construction 
of the quantum analogs of (\ref{mathHcl}) and 
(\ref{Q+clas}) do not cure the `disease' of the quantum 
anomaly due to the quadratic in $\hbar$ nature of the 
necessary `remedy' in the form of the 
correction terms both in the Hamiltonian and in the supercharges.


\vskip0.1cm

So, let us consider a pair of the quantum systems 
described by the Hamiltonian operators $\hat{H}_\pm$, 
\be\label{h+-}
	2\hat{H}_\pm=-\frac{d^2}{dx^2}+V_\pm(x)\,,
\ee
and require that they are intertwined,
\be\label{intert}
	\hat{q}_+\hat{H}_-=\hat{H}_+\hat{q}_+\,,
\ee
 by the second order differential operator $\hat{q}_+$,
\be\label{q+}
	2\hat{q}_+=\frac{d^2}{dx^2}+
	s_1(x)\frac{d}{dx}+s_0(x)\,.
\ee
Then they also will be  intertwined  by $\hat{q}_-=(\hat{q}_+)^\dagger$
in the opposite direction:
$\hat{q}_-\hat{H}_+=\hat{H}_-\hat{q}_-$.
In (\ref{h+-}) and (\ref{q+}) we set $\hbar=1$, but later we shall reconstruct 
Planck constant. 
{}Operators $\hat{H}_\pm$  and 
$\hat{q}_\pm$ represent the composing blocks for
the supersymmetric Hamiltonian $\hat{\mathcal{H}}=\text{diag}\,(\hat{H}_+,\hat{H}_-)$ and  
the second order  
supercharges  $\hat{\mathcal{Q}}_+=\hat{q}_+\sigma_+$,
 $\hat{\mathcal{Q}}_-=\hat{q}_-\sigma_-=\hat{\mathcal{Q}}_+^\dagger$.

The intertwining condition (\ref{intert}) 
produces three equations which appear from equating 
the coefficients at
$\frac{d^2}{dx^2}$, $\frac{d}{dx}$ and $1$:
\be\label{3equal}
	V_-=-2s_1'+V_+\,,\quad
	 2V'_-+s_1V_-=-s_1''-2s_0'+s_1V_+\,,\quad
	V_-''+s_1V'_-+
	s_0V_-=-s_0''+V_+s_0\,.
\ee
The first equation gives 
\be\label{V+-}
	V_+=V_-+2s_1'\,.
\ee
Substitution of (\ref{V+-}) into the second equation yields  
\be\label{V+-s01}
	V_-=\frac{1}{2}s_1^2-\frac{1}{2}s_1'-s_0+2a\,,
\ee
where $a$ is an integration  constant. Substitution of (\ref{V+-})
and (\ref{V+-s01}) into the third equation 
from (\ref{3equal}) allows us to 
express  $s_0(x)$ in terms of $s_1(x)$,
\be\label{s0(x)}
	s_0=\frac{1}{4}s_1^2+\frac{1}{2}s_1'-\frac{1}{2}\frac{s_1''}{s_1}+\frac{1}{4}
	\left(\frac{s_1'}{s_1}\right)^2+\frac{\tilde{C}}{s_1^2}\,,
\ee
where $\tilde{C}$ is an integration  constant.
Substituting (\ref{s0(x)}) into (\ref{V+-s01}) and introducing the notation
\be\label{Ws1}
	W(x)=\frac{1}{2}s_1(x)\,,
\ee
we present the potential $V_-(x)$ in the form 
\be\label{V-(W)}
	V_-=w_-^2-w_-'-\frac{C}{W^2}+2a\,,\qquad
	w_-\equiv W-\frac{1}{2} \frac{W'}{W}\,,
\ee
where $C=\tilde{C}/4$.
Using Eq.  (\ref{V+-}), we fix then  $V_+(x)=V_-(x)+4W'(x)$ in terms of 
$W(x)$ in a  form similar to (\ref{V-(W)}),
 \be\label{V+(W)}
	V_+=w_+^2+w_+'-\frac{C}{W^2}+2a\,,\qquad
	w_+\equiv W+\frac{1}{2} \frac{W'}{W}\,.
\ee
Note that $w_+(W)=-w_-(-W)$, $V_+(W)=V_-(-W)$.
Function $s_0(x)$ from (\ref{s0(x)}) can be presented in one of the forms
\be\label{s0(x)+}
	s_0=w_-'-w_-^2+\frac{C}{W^2}+2Ww_-=-w_+'-w_+^2+\frac{C}{W^2}+2Ww_++2W'\,.
\ee
The intertwining operators $\hat{q}_+$ and $\hat{q}_-=\hat{q}_+^\dagger$
are given finally  in terms of the superpotential  $W(x)$, 
and with the help of 
(\ref{s0(x)+}) we find   that $\hat{q}_+(-W)=\hat{q}_-(W)$.
This relation together with the relation $V_+(W)=V_-(-W)$ corresponds to classical
symmetry transformation (\ref{classymd}).

At first glance it would seem that 
one has to require that  function (\ref{Ws1}) must be a never vanishing 
function in order the potentials $V_-(x)$ and $V_+(x)$ to be 
well defined (regular) functions. However, $W(x)$ enters into $V_-(x)$ and $V_+(x)$
in a nontrivial  way, see Eq. (\ref{V-(W)}), (\ref{V+(W)}). As a consequence,
as it will be shown by specific examples in Section \ref{Examples},  
for some choices of  $W(x)$ having zeros on the real line,
one can obtain the physically interesting super-partner  potentials 
$V_\pm(x)$ to be regular on all the real line by 
the appropriate choice of the constant $C$. On the other hand,
to specify  the  operators appearing in the constructions,
it is necessary to declare their corresponding domains.
We just will  have this in mind but without making an explicit  specification
of the domains in order do not overload the presentation 
with  technical details.

Planck constant $\hbar$  is reconstructed by employing 
 the mnemonic rule
in (\ref{h+-}), (\ref{q+}):
each time when there appears the operator $\frac{d}{dx}$ or  derivative $W'$, they should be 
multiplied by $\hbar$, while  $\frac{d^2}{dx^2}$ and $W''$ should be 
accompanied by  the factor
$\hbar^2$.  For the Hamiltonian superpartner operators $\hat{H}_\pm$ we then have
\be\label{H+-AA}
	2(\hat{H}_--a)=A^\dagger_-A_- -\frac{C}{W^2}\,,\qquad
	2(\hat{H}_+-a)=A_+ A_+^\dagger  -\frac{C}{W^2}\,,
\ee
where 
\be\label{w+-def} 
	A_-=
	\hbar\frac{d}{dx}+w_-
	\,,\qquad
	A_+
	=
	\hbar\frac{d}{dx}+w_+
	\,,\qquad
	w_\pm=W\pm \hbar \frac{W'}{2W}\,.
\ee
The intertwining operator  $\hat{q}_+$  can be presented in the form 
\be\label{q+AA}
	2\hat{q}_+=A_+A_-+\frac{C}{W^2}\,.
\ee
In terms of the first order operators  $A_\pm$, the  
validity of intertwining relations  (\ref{intert}) and 
the conjugate intertwining relations  for $\hat{q}_-$ 
can be checked  using the 
identities
\begin{eqnarray}
	&A_-A_-^\dagger=A^\dagger_+ A_+\,,&\label{id1}\\
	&A_+ + A_-^\dagger=2W\,,\qquad 
	A_- + A_+^\dagger=2W\,,&\label{id2}\\
	&\frac{1}{W}A_-=A_+\frac{1}{W}\,,\qquad
	\frac{1}{W}A_+^\dagger=A_-^\dagger \frac{1}{W}\,.&\label{id3}
\end{eqnarray}
Relations  (\ref{id3}) show  that   function
$\frac{1}{W}$ is a kind of the  intertwining operator 
for the pairs of the first order differential operators 
$A_-$ and $A_+$, and $A_+^\dagger$ and $A_-^\dagger$.

 \vskip0.2cm
 In accordance with relations (\ref{H+-AA}) and 
 (\ref{q+AA}), 
 in the case of $C=0$ the operators $A_\pm$
and the Hermitian conjugate operators 
 $A^\dagger_\pm$ provide a factorization 
of the Hamiltonians $\hat{H}_-$ and $\hat{H}_+$ (after shifting them for $-a$),
as well as of the  second order  intertwining operators $\hat{q}_+$ and $\hat{q}_-$.
This means particularly 
that the  intertwining between $\hat{H}_-$ and $\hat{H}_+$ 
can be  realized in this case in two steps through the 
 intermediate, virtual  Hamiltonian operator 
 $\hat{H}_{{}_\otimes}$, which is  given by 
 $2(\hat{H}_{{}_\otimes}-a)\equiv
{A}_-{A}_-^\dagger={A}_+^\dagger {A}_+$,
see Eq.  (\ref{id1}), 
	and satisfies  the relations 
${A}_-\hat{H}_-=\hat{H}_{{}_\otimes} {A}_-$,
${A}_+\hat{H}_{{}_\otimes}=\hat{H}_ +{A}_+$.
 
The  operators  $\hat{H}_-$,  $\hat{H}_+$, 
$\hat{q}_+$ and $\hat{q}_-$ can be presented 
in the factorized form also in the case of  $C\neq 0$.
To find such a factorization together with the intermediate, 
virtual  Hamiltonian operator appearing in the 
intertwining relations,  introduce the first order differential operators 
\be\label{calA}
	\mathcal{A}_\pm=A_\pm\pm \lambda(x)\,,
	\qquad \mathcal{A}^\sharp_\pm=A^\dagger_\pm\pm \lambda(x)
\ee
with  some function $\lambda(x)$, and require that 
the shifted Hamiltonian operators (\ref{H+-AA}) 
are presented in 
the equivalent form 
\be\label{intert1}
	2(\hat{H}_- - b)=\mathcal{A}^\sharp_-\mathcal{A}_-\,,\qquad
	2(\hat{H}_+ - b)-\alpha=\mathcal{A}_+\mathcal{A}^\sharp_+\,,
\ee
where $b$ and $\alpha$ are some constants.
One  finds  that the indicated equivalence  takes place 
if 
\be\label{abalp}
	\lambda(x)=-\frac{\alpha}{4W(x)}\,,\qquad
	b=a-\frac{1}{4}\alpha\,,\qquad
	\alpha^2=-16C\,.
\ee
As a consequence of  (\ref{intert1}), 
the intertwining  operators are 
presented in the equivalent form
\be \label{intert2}
 	2\hat{q}_+=\mathcal{A}_+\mathcal{A}_-\,,\qquad
	2\hat{q}_-=\mathcal{A}_-^\sharp \mathcal{A}_+^\sharp\,,
\ee
and we also have a relation
\be\label{intert3}
	\mathcal{A}_-\mathcal{A}_-^\sharp=\mathcal{A}_+^\sharp \mathcal{A}_+
	+\alpha\,,
\ee
cf. (\ref{id1}).
When $C=-\beta^2<0$, parameter 
$\alpha$ is fixed to be real and can take any of
two values $\alpha=\pm 4\beta$,
$\beta\in\R$, $\beta\neq 0 $. 
In this case $\mathcal{A}^\sharp_\pm=\mathcal{A}^\dagger_\pm$,
and the intermediate Hamiltonian $H_{{}_\otimes}$,
which is  given now by relations
 $2(\hat{H}_{{}_\otimes}-b)\equiv
\mathcal{A}_-\mathcal{A}_-^\sharp=\mathcal{A}_+^\sharp \mathcal{A}_+
	+\alpha$ 
	and is subject to  the intertwining relations 
$\mathcal{A}_-\hat{H}_-=\hat{H}_{{}_\otimes} \mathcal{A}_-$,
$\mathcal{A}_+\hat{H}_{{}_\otimes}=\hat{H}_ +\mathcal{A}_+$,
is Hermitian.
On the other hand, when 
$C=\kappa^2>0$, we have $\alpha=\pm 4i\kappa$,
$\kappa\in\R$, $\kappa\neq 0 $. 
In this case $\mathcal{A}^\sharp_\pm\neq
\mathcal{A}^\dagger_\pm$, the intermediate 
 Hamiltonian $H_{{}_\otimes}$ is not Hermitian,
 and the shift parameter $b$ in (\ref{intert1})
 is complex, $b^*\neq b$.
 Notice that
 the factorization (\ref{intert1})  in the case of nonzero value of the 
 parameter $C$ requires different constant 
 shifts  in supersymmetric partner Hamiltonians 
 $\hat{H}_-$ and $\hat{H}_+$.
 These shifts as well as the shift  in (\ref{intert3}) 
 are different  for different signs of the parameter $\alpha$.
 The peculiarities associated with 
 the change of sign of the parameter $\alpha$ 
 are illustrated by some examples 
 in Section \ref{Examples}.  
 Let us stress also here that fixing 
 parameter $\alpha$ does not 
 specify yet uniquely the 
 operators $\mathcal{A}_\pm$ and 
 $\mathcal{A}_\pm^\dagger$.
 This point is discussed in Appendix.
 
Using relations (\ref{intert1}),  (\ref{intert2}),
 (\ref{intert3}) and (\ref{abalp}), we find that 
the Hamiltonian $\hat{\mathcal{H}}$ of the system, 
composed from superpartners $\hat{H}_-$ and
$\hat{H}_+$, and supercharges $\hat{\mathcal{Q}}_+$
and $\hat{\mathcal{Q}}_-=\hat{\mathcal{Q}}_+^\dagger$,
\be\label{HqQpmq}
	\hat{\mathcal{H}}=
	\left(
	\begin{array}{cc}
	 \hat{H}_+ &   0   \\
	0    &     \hat{H}_-
	\end{array}
	\right),\qquad
	\hat{\mathcal{Q}}_+=
	\left(
	\begin{array}{cc}
	0 &   \hat{q}_+   \\
	0    &    0
	\end{array}
	\right),\qquad
	\hat{\mathcal{Q}}_-=
	\left(
	\begin{array}{cc}
	0 &  0   \\
	 \hat{q}_-    &    0
	\end{array}
	\right),\qquad
\ee
satisfy the
relations $[\hat{\mathcal{H}},\hat{\mathcal{Q}}_\pm]=0$,
$\hat{\mathcal{Q}}_+^2=
\hat{\mathcal{Q}}_-^2=0$, and 
\be\label{QQquant}
	[\hat{\mathcal{Q}}_+,\hat{\mathcal{Q}}_-]_+=
	(\hat{\mathcal{H}}-a)^2+C\,,
\ee
 cf.
(\ref{QQHclas}).	

\vskip0.2cm

The superpartner Hamiltonians  (\ref{H+-AA}) and the intertwining 
operator (\ref{q+AA})   can be presented  in  the  form 
\be\label{HWh}
	2(\hat{H}_\mp-a)=-\hbar^2\frac{d^2}{dx^2}+W^2\mp 2\hbar W'-\frac{C}{W^2}+\Delta(W)\,,
\ee
\be\label{q+quant}
	2\hat{q}_+=\left(\hbar\frac{d}{dx}+W\right)^2+\frac{C}{W^2}-\Delta(W)\,,
\ee
where 
\be\label{Delta(W)}
	\Delta(W)=\frac{1}{2}\hbar^2\left(\frac{W''}{W}-\frac{1}{2}\left(\frac{W'}{W}\right)^2\right)=
	\hbar^2\frac{1}{\sqrt{W}}\left(\sqrt{W}\right)''=
	\hbar^2(\omega^2+\omega')\,,
\ee
and 
$
\omega= \frac{W'}{2W}\,. 
$
The mass parameter can  be
restored  by changing the quantum factor $\hbar$ in relations (\ref{w+-def} ),
(\ref{HWh}), (\ref{q+quant})  and 
(\ref{Delta(W)}) for 
$\frac{\hbar}{\sqrt{m}}$.

Note  here that 
a similar   term quadratic in Planck constant $\hbar^2$ emerges  
in the quantum Hamilton-Jacobi equation in the form of the 
quantum potential $V_Q=-\frac{\hbar^2}{2m}\frac{1}{\sqrt{\rho}}(\sqrt{\rho})''$,
where $\rho$ is the probability density \cite{QuantHJ}.
Introduce  a    function $z(x)$ given locally by a relation
\be\label{x->z}
	z'(x)=\frac{1}{W(x)}\,.
\ee
Reconstructing the mass parameter,
Eq.  (\ref{Delta(W)}) can be written then 
in the equivalent form 
\be\label{DeltaSchwarz} 
	\Delta(x)=-\frac{\hbar^2}{2m} S(z(x))\,,
\ee
 where 
 \be
 S(z(x))\equiv \left(\frac{z''}{z'}\right)'-\frac{1}{2}\left(\frac{z''}{z'}\right)^2=
 \frac{z'''}{z'}-\frac{3}{2}\left(\frac{z''}{z'}\right)^2
 \ee
 is the Schwarzian derivative, which is a third-order differential 
 expression. 
 It is interesting to compare (\ref{DeltaSchwarz} )  
 with a  presentation  for a shifted potential 
 $u(x)-E$  in 
 the stationary Schr\"odinger equation 
 $-\frac{\hbar^2}{2m}\psi''(x)+(u(x)-E)\psi(x)=0$
 when it is expressed in terms of an arbitrary 
 solution in the form 
 \be\label{Schf}
 u(x)-E=-\frac{\hbar^2}{2m} S(f(x))\,, 
 \ee
 where $f(x)=\int^x \frac{ds}{\psi^2(s)}$, cf. (\ref{x->z}). 
 Function  $f(x)$  is not fixed uniquely here and 
 can be changed for a quotient $f(x)=\frac{\psi_1(x)}{\psi_2(x)}$ 
 of two arbitrary linearly independent solutions of the 
 Schr\"odinger equation. This  reflects 
 one of the basic properties of the 
 Schwarzian derivative : it is  invariant under 
 arbitrary M\"obius transformation   $f(x)\rightarrow F(x)=\frac{a f(x)+b}{cf(x)+d}$
 of the  functional argument, $S(F(x))=S(f(x))$. 
 Note also here the relations
\be
S(\phi(x))=S(\psi(x))+S(f(\psi))\psi'\,,\qquad
S(\phi(x))=h'{}^2S(\phi(y))+S(h(x))\,,
\ee
which correspond to arbitrary changes of dependent and independent variables,
respectively,
where $\phi=f(\psi)$,  $y=h(x)$,
$\psi'=d\psi/dx$, $h'=dh/dx$.

If we compare now the quantum Hamiltonian and the supercharges 
given by (\ref{HqQpmq}), (\ref{HWh}) and  (\ref{q+quant}), 
with the operators $\hat{\mathcal{H}}_{d}$ and $\hat{\mathcal{Q}}_{+d}$
which we obtained by quantization of the classical system, 
we  find that  
\be\label{quantanom}
	\hat{\mathcal{H}}-\hat{\mathcal{H}}_{d}=\frac{1}{2}\Delta(W)\,,\qquad
	\hat{\mathcal{Q}}_+-\hat{\mathcal{Q}}_{+d}=-\frac{1}{2}\Delta(W)\hat{\theta}_+\,.
\ee
The difference is given  by the  same 
quadratic in $\hbar^2$ quantum term $\Delta(W)$,
which, as it follows from (\ref{Delta(W)}),
  turns into zero only for quadratic 
superpotential of the form $W(x)=(a_1x+a_0)^2$. 
Only in this case the direct quantization 
prescription we discussed above 
yields the Hamiltonian and supercharge
operators coinciding exactly with 
the quantum Hamiltonian $\hat{\mathcal{H}}$ 
and corresponding second order supercharges,
and the quantum anomaly is absent. 
This reflects another but related  basic property 
of the Schwarzian derivative. Namely, 
$S(z(x))=0$ if and only if $z(x)$ 
is a fractional linear transformation
$z(x)=\frac{ax+b}{cx+d}$.  
For such a function we have   $z'(x)=\frac{ad-bc}{(cx+d)^2}$, and
in accordance with  (\ref{x->z}) this corresponds to
$W(x)=(a_1 x+a_0)^2$.

To find the anomaly-free quantization prescription
in the case of arbitrary superpotential $W(x)$, 
we note that the factor $\hbar^2$ is present also in the 
quantum  kinetic  term $-\hbar^2\frac{d^2}{dx^2}$ 
in (\ref{HWh}) as well as in 
(\ref{q+quant}).   
Returning to the classical level,  the   
kinetic term $p^2$ can be rewritten
in an equivalent form 
\be\label{p2equiv}
	\zeta p\frac{1}{\zeta^2} p \zeta=
	\left(\zeta(-i p)\frac{1}{\zeta}\right)\left(\frac{1}{\zeta} ip \zeta\right)
\ee
involving an  arbitrary function $\zeta=\zeta(x)$.
 Taking  $\zeta(x)=\frac{1}{\sqrt{W(x)}}$, as a direct quantum analog
 of the classical factor $\frac{1}{\zeta(x)} ip \zeta(x)$ 
we obtain 
$\hat{\mathcal{P}}=\hbar(\frac{d}{dx}-\frac{W'}{2W})$,  and 
$\hat{\mathcal{P}}^\dagger =-\hbar(\frac{d}{dx}+\frac{W'}{2W})$ 
will be the quantum analog of the factor
$\zeta(x) (-ip)\frac{1}{\zeta(x)}$. 
As the quantum analog of 
$p^2=\left(\zeta (-ip)\frac{1}{\zeta}\right) \left(\frac{1}{\zeta }ip \zeta \right)$ 
we take then
$\hat{\mathcal{P}}^\dagger \hat{\mathcal{P}}$. 
This prescription gives 
the operator
$-\hbar^2\frac{d^2}{dx^2} +\Delta(x) +W^2$
as the quantum analog of the 
classical term $zz^*=p^2+W^2$ 
from the Hamiltonian (\ref{mathHcl}).  
Taking as before $\hat{N}=\frac{1}{2}\hbar\sigma_3$,  
we find that
the quantum analog of $\mathcal{H}$ 
will coincide as a result  with $\hat{\mathcal{H}}$.
In analogous way  
we rewrite  the term $z^2$ from 
supercharge  (\ref{Q+clas}) 
in the classically equivalent form 
$(ip+W)^2=\left(\zeta z \frac{1}{\zeta}\right)
\left(\frac{1}{\zeta}z\zeta\right)$,
and the  quantum analog of this will give us the operator
$(\hbar\frac{d}{dx}+w_+)(\hbar\frac{d}{dx}+w_-)=A_+A_-$,
see  (\ref{w+-def}).
The quantum analog of $Q_+$ will reproduce then exactly $\hat{Q}_+$~\footnote{
The manipulations we realize 
at the quantum level, which are  associated with the classical
change (\ref{p2equiv}),  were discussed in more detail 
in \cite{BMP}.  We only  stress here that  the 
final quantum change $\hat{\mathcal{H}}_{d}\rightarrow \hat{\mathcal{H}}$,
$\hat{\mathcal{Q}}_{\pm d}\rightarrow \hat{\mathcal{Q}}_{\pm}$ we 
obtain is, obviously,  not a unitary transformation :
$[\hat{\mathcal{H}}_{d},\hat{\mathcal{Q}}_{\pm d}]\neq 0$ while
$[ \hat{\mathcal{H}}, \hat{\mathcal{Q}}_{\pm}]=0$. }.

The indicated fictitious similarity transformation
in the classical  kinetic term was exploited recently 
in \cite{BMP} in the  discussion of the quantum systems
with position-dependent mass. 
It is similar to a structure appearing in
the  problem of quantization of  a particle in a curved metric
\cite{deWitt},
and was also used  in solving the problem of
the operator ordering ambiguity in 
supersymmetric quantum mechanics 
with linear supercharges in curved space 
\cite{deAlf}.
In the present case we are 
in a space of one dimension and so, the space  is 
flat~\footnote{Though 
curvature is irrelevant aspect in one-dimensional cases, 
the topology of the configuration space may be relevant in some 
systems.}. 
We already noted a similarity of the specific 
quadratic in $\hbar$ quantum 
term (\ref{Delta(W)}) with the quantum potential term 
$V_Q$ appearing  in the 
quantum Hamilton-Jacobi equation.
Interestingly,  
in the case of multi-dimensional 
Euclidean space the  $V_Q$ term  can be treated  
as a curvature of the 
amplitude of the wave function \cite{QuantHJ}. 
It is worth to note that 
the Schwarzian derivative
admits also the interpretation 
in terms of the curvature if we consider a free 
particle motion in the 
Lorentz plane  and reduce it to a given curve, 
see \cite{DuvOvs,OvsTab} and  \cite{BMP}. 
The Schwarzian derivative appears also 
in the Schr\"odinger equation under application 
to the latter of a form-preserving change of variable and 
wave function, see Section \ref{SchwSchr}.
For some other applications of the Schwarzian derivative
and its properties see also 
refs. \cite{Chu,Osg,Gib,Carin,Toda,Weiss,Chaos}. 

\section{Coupling-constant metamorphosis and duality}\label{Metamor}
 Before we consider some examples, 
 let us discuss the general picture 
 we obtained in the related interesting 
 context of the coupling-constant 
 metamorphosis \cite{Hieat,HietGram,CabGal,Cariglia}. 
 To this aim, consider the stationary Schr\"odinger equations 
 corresponding to the quantum system (\ref{HqQpmq}) 
 characterized by the second order supercharges,
 \be\label{H+-E}
 	(\hat{H}_\mp -E)\Psi_\mp(x)=0\,.
 \ee
 We set  $\Psi_\mp(x)=\sqrt{W(x)}\psi_\mp(x)$,
 multiply the equation from the left by $W^2(x)$,
 change the variable $x$ for $z$ 
 using relation (\ref{x->z}), $z'=1/W(x)$,
and
denote $\widetilde{W}(z)=W(x(z))$, 
 $\widetilde{\psi}_\mp(z)=\psi_\mp(x(z))$.
 In this way  Eq. (\ref{H+-E}) 
 will be transformed  into 
 \be\label{Htransf}
	\left(-\hbar^2 \frac{d^2}{dz^2} + \widetilde{W}^4
	\mp\hbar \frac{d\widetilde{W}^2}{dz}-C-\widetilde{W}^2E\right)
	\widetilde{\psi}_\mp(z)=0\,,
\ee
where we included the additive constant $-a$ from (\ref{HWh})  into $E$.
Introducing the notation 
$\mathcal{W}=\widetilde{W}^2-\frac{1}{2}E$,
equation (\ref{H+-E}) takes finally
the form  
\be\label{susy1}
	\left(-\hbar^2 \frac{d^2}{dz^2} + \mathcal{W}^2
	\mp\hbar \frac{d\mathcal{W}}{dz}-\mathcal{E}\right)
	\widetilde{\psi}_\mp(z)=0\,,
\ee
where $\mathcal{E}=C+\frac{1}{4}E^2$.
We have arrived at 
a pair of the stationary 
Schr\"odinger equations corresponding to a supersymmetric system
characterized by the linear supercharges
$\hat{\mathcal{S}}_+=(\mathcal{W}+\hbar\frac{d}{dz})\sigma_+$
and $\hat{\mathcal{S}}_-=\hat{\mathcal{S}}_+^\dagger$.
The pair of the superpartner potentials $U_\mp(z)=
\mathcal{W}^2
\mp\hbar \frac{d\mathcal{W}}{dz}
$
is given in terms of the superpotential $\mathcal{W}(z)$.
We have here a kind of a metamorphosis  \cite{Hieat},
which exchanges the roles of the
coupling constant and the energy 
in the superpartner systems, 
and the associated  specific duality 
between 
the systems with the second (\ref{H+-E})  and first order (\ref{susy1}) supersymmetries.
The peculiarity of the coupling-constant metamorphosis 
appearing here is 
that  the combination $C+\frac{1}{4}
E^2$ playing a role of the energy eigenvalue $\mathcal{E}$
depends not only on the constant $C$, but also on 
the energy eigenvalue $E$
of the initial superpartner systems.
It is necessary to stress, however,  that 
the described metamorphosis and the associated 
transmutation of the second order supersymmetry into the 
supersymmetry with the first  order supercharges 
was obtained here formally, by employing relation (\ref{x->z}) 
to define the function $x(z)$ locally,
and we are not preoccupied with the 
global nature of the function $W(x)$.

Classically, the described metamorphosis of the coupling constant 
and transmutation of
the order of supersymmetry can be understood in correspondence
with the picture of duality between integrable Hamiltonian
systems discussed by Hietarinta et al \cite{Hieat}. 
To this end, we multiply classical Hamiltonian 
(\ref{mathHcl}) by $W^2$. After changing 
$2(\mathcal{H}-a)$ for a constant $\eta$, 
and $C+\frac{1}{4}\eta^2$ for $2\mathcal{G}$, 
and making a point canonical transformation
$x\rightarrow z$, $p\rightarrow P$, where $z(x)$ is given by
Eq. (\ref{x->z}) and $P=W(x)p$, we
arrive at the classical system described by the Hamiltonian
$\mathcal{G}$ given by
\be\label{clasmetamor}
	2\mathcal{G}=P^2+\mathcal{W}^2+2\frac{d\mathcal{W}}{dz}N\,,
\ee
cf. (\ref{susy1}), where $\mathcal{W}(z)=\widetilde{W}^2(z)-\frac{1}{2}\eta$,
$\widetilde{W}(z)=W(x(z))$.
In correspondence with the quantum picture,
the classical system is characterized by the supercharges 
$\mathcal{S}_+=(\mathcal{W}(z)+iP)\theta^+$ and $\mathcal{S}_-=(\mathcal{S}_+)^*$,
which are the first order integrals  in the momentum $P$,
$\{\mathcal{G},\mathcal{S}_\pm\}=0$, 
$\{\mathcal{S}_+,\mathcal{S}_-\}=-2i\mathcal{G}$.
The first order  supercharges $\mathcal{S}_\pm$ of the system $\mathcal{G}$ 
can be obtained 
from the second order supercharges $\mathcal{Q}_\pm$
of the system $\mathcal{H}$. 
To see this, it is sufficient to note that $\mathcal{Q}_+$ can be rewritten 
in the equivalent form $\mathcal{Q}_+=\left(-2(\mathcal{H}-a)+2(W^2+ipW)\right)\theta^+$.
After the change $2(\mathcal{H}-a)\rightarrow\eta$, and realization
of the indicated point canonical transformation,  $\mathcal{Q}_+$ 
will transform into $2\mathcal{S}_+$.
Corresponding supercharge operator $\hat{\mathcal{S}}_+$
can also be obtained from $\hat{\mathcal{Q}}_+$
by a procedure similar to that how the 
superpartner pair of Hamiltonians in  (\ref{susy1}) was obtained.

 \section{Schr\"odinger equation and Schwarzian derivative}\label{SchwSchr}

To understand better the roots of the Schwarzian derivative 
`remedy' for the quantum anomaly `disease' 
that we employed in Section \ref{AnomalySec}
as well as to see a  hidden role played by the Schwarzian  
in  the coupling-constant  metamorphosis mechanism discussed in 
the previous Section,  
we  consider here the form-invariant transformations
in stationary Schr\"odinger equation.

So, let us take the stationary Schr\"odinger equation
\be\label{SEx}
	\left(-\frac{\hbar^2}{2m}\frac{d^2}{d\xi^2}+V(\xi)-E\right)\psi_E(\xi)=0\,,
\ee
and change in it the variable $\xi$ for some another  variable $x$ 
given by the relation 
$\frac{dx}{d\xi}=f(\xi)$. 
Then
$
d\xi=\frac{dx}{\varphi(x)}\,,
$
where $\varphi(x)=f(\xi(x))$,
cf. (\ref{x->z}), and 
$\frac{d}{d\xi}=\varphi(x)\frac{d}{dx}$, 
$\frac{d^2}{d\xi^2}=\varphi^2\frac{d^2}{dx^2}+\varphi\varphi'\frac{d}{dx}$. 
After the change of variable and multiplication by $\frac{1}{\varphi^2(x)}$
from the left, Eq. 
(\ref{SEx}) transforms into
\be\label{SEz}
	-\frac{\hbar^2}{2m}\left(\frac{d^2}{dx^2}+\frac{\varphi'}{\varphi}\frac{d}{dx}
	\right)\widetilde{\psi}_E+
	\frac{\tilde{V}(x)-E}{\varphi^2}\widetilde{\psi}_E(x)=0\,,
\ee
where $\widetilde{V}(x)=V(\xi(x))$, $\widetilde{\psi}_E(x)=\psi_E(\xi(x))$.
The first order derivative term is eliminated 
by defining  a new function $\breve{\psi}_E(x)$, 
$
\widetilde{\psi}_E(x)=\varphi^{-1/2}(x)\breve{\psi}_E(x)\,,
$
for which Eq.  (\ref{SEz}) transforms into
\be\label{SEbre}
	\left(-\frac{\hbar^2}{2m}\frac{d^2}{dx^2}+\frac{1}{2}\Delta(\varphi)+
	\frac{\tilde{V}(x)-E}{\varphi^2}\right)\breve{\psi}_E(x)=0,
\ee
 where 
 $$
 \Delta(\varphi)=\frac{\hbar^2}{2m}
 \left(\frac{\varphi''}{\varphi}-\frac{1}{2}\left(\frac{\varphi'}{\varphi}\right)^2\right)=
 \frac{\hbar^2}{m}\left(\left(\frac{\varphi'}{2\varphi}\right)'+\left(\frac{\varphi'}{2\varphi}\right)^2\right)\,,
 $$
 cf. (\ref{Delta(W)}). The appearance of the $\Delta$ term here is 
 associated ultimately with the identity relations 
 \be\label{useindent}
 	f^{-3/2}(\xi)\frac{d^2}{d\xi^2}f^{-1/2}(\xi)=\varphi^{-1/2}\frac{d}{dx}\varphi\frac{d}{dx}\varphi^{-1/2}=
 	\frac{d^2}{dx^2}
	 -\frac{1}{\sqrt{\varphi}}(\sqrt{\varphi})''\,.
 \ee 
 Remembering that $\varphi(\xi)=1/\xi'(x)$, the last  term 
 in (\ref{useindent}) corresponds to the Schwarzian derivative:
 $-\frac{1}{\sqrt{\varphi}}(\sqrt{\varphi})''=\frac{1}{2}S(\xi(x))$,
 $S(\xi(x))=\frac{\xi'''}{\xi'}-\frac{3}{2}\left(
\frac{\xi''}{\xi'}\right)^2$. 
 
 In the discussion of the coupling-constant metamorphosis 
 we realized a procedure that, in fact, is inverse 
 to the procedure 
 described here and which corresponds to a transformation
 from (\ref{SEbre}) to (\ref{SEx}). 
 The middle 
differential term in (\ref{useindent}) can be compared 
with the representation (\ref{p2equiv})  for the kinetic 
term  we used to formulate the anomaly-free prescription for
quantization of the classical system.


\section{Some examples}\label{Examples}
To illustrate the obtained general results, we consider here several   
simple examples. 

$\bullet$ For  $W=\gamma x$,  we have 
 $\Delta=-\frac{\hbar^2}{4}\frac{1}{x^2}<0$,
 and 
 $V_\mp=\gamma^2x^2-\left(\frac{\hbar^2}{4}+\frac{C}{\gamma^2}\right)
 \frac{1}{x^2}\mp 2\hbar \gamma$.
With the choice   $C=-\frac{\hbar^2\gamma^2}{4}<0$  both potentials 
$V_\mp(x) = \gamma^2 x^2 \mp 2\hbar \gamma$ are regular on $\R$ functions 
which correspond to  a pair of 
harmonic oscillators with mutually shifted spectra $E^{-}_n=\hbar\gamma n$,
$E^{+}_n=\hbar\gamma (n+2)$, $n=0,1,\ldots$, where we set additive 
constant $a=0$.  We have here $w_\pm=\gamma x\pm\frac{\hbar}{2x}$,
and so, the first order operators $A_\pm$ given by Eq. (\ref{w+-def} ) are 
singular at $x=0$.
With $\alpha=2\hbar\gamma$ and $\lambda(x)=-\frac{\hbar}{2x}$,
see Eq. (\ref{abalp}), 
we find that  operators $\mathcal{A}_\pm$ are equal and coincide 
(up to inessential constant factor) with the annihilation harmonic oscillator operator:
$\mathcal{A}_-=\mathcal{A}_+={\hbar}\frac{d}{dx}+\gamma x$.
The superpartner Hamiltonians are represented in terms of these and conjugate operators
as $2\hat{H}_-=\mathcal{A}_-^\dagger\mathcal{A}_-$,
$2\hat{H}_+=\mathcal{A}_-\mathcal{A}_-^\dagger+2\hbar\gamma$.
They are intertwined by $2\hat{q}_+=\mathcal{A}_-^2=
\left({\hbar}\frac{d}{dx}+\gamma x\right)^2$.
The intermediate, virtual system $H_\otimes$ is given by
$2H_\otimes=\mathcal{A}_-\mathcal{A}_-^\dagger=
2\hat{H}_-+\hbar\gamma$ that corresponds to the harmonic 
oscillator with potential minimum shifted exactly to the middle 
between the minima of potentials $V_-(x)$ and $V_+(x)$. 

For the alternative  choice $\alpha=-2\hbar\gamma$, we
have $\lambda(x)=+\frac{\hbar}{2x}$, and then 
the factorizing operators $\mathcal{A}_\mp$ are different and singular:
$\mathcal{A}_\mp={\hbar}\frac{d}{dx}+\gamma x\mp\frac{\hbar}{x}$.
In this case in correspondence with relation 
$\mathcal{A}_-\mathcal{A}_-^\sharp=-\hbar^2\frac{d^2}{dx^2}
+\gamma^2x^2+2\frac{\hbar^2}{x^2}-\gamma\hbar$,
 the same mutually shifted superpartner 
 harmonic oscillators  
 given by potentials 
  $V_\mp(x) = \gamma^2 x^2 \mp 2\hbar \gamma$ are intertwined
 through the isotonic oscillator system,
 see  \cite{BerUss1,iso1,iso2,iso3}. 
 It could seem that there is no sense
 to consider such alternative singular (at $x=0$ here) factorizing operators  
 for regular on all the real line $\R$  superpartner Hamiltonians.
 Nevertheless, the alternative singular factorizing operators
 had prove to be very important  for instance under analysis of the 
 phenomenon of supersymmetry transmutations 
 in exotic supersymmetric structure associated with 
 soliton (reflectionless) systems,
 see refs. \cite{AGP1,AnMP2}.

If  $x$ is treated as a radial variable, 
then with the  choice of $C<-\frac{\hbar^2\gamma^2}{4}$
one can relate the picture  to a supersymmetric 
pair of the  $3D$ isotropic oscillators \cite{CKSb}.

In the context of 
the coupling-constant metamorphosis mechanism,
as a dual supersymmetric system 
with the  first order supercharges we obtain   the system with a pair of
Morse potentials of the form $U_\mp(z)=\gamma^4e^{4\gamma z}-2\gamma^2
(\kappa\pm \hbar\gamma)e^{2\gamma z}+\kappa^2$
given in terms of a superpotential  
$\mathcal{W}=\gamma^2e^{2\gamma z}-\kappa$.
For the discussion of such a supersymetric system see, e.g. \cite{Cooper,CKSb}.

\vskip0.1cm

$\bullet$ For  $W=\frac{\gamma}{x}$, we find  $\Delta=\frac{3}{4}\hbar^2 \frac{1}{x^2}>0$,
$V_\mp=\frac{1}{x^2}((\gamma\pm \hbar)^2-\frac{1}{4}\hbar^2)-\frac{C}{\gamma^2} x^2$,
$w_\mp=(\gamma\pm\frac{\hbar}{2})\frac{1}{x}$.
The choice  $\gamma=\frac{1}{2}\hbar$ or $\gamma=\frac{3}{2}\hbar$ 
gives $V_+=-\frac{C}{\gamma^2}x^2$ and $V_-=-\frac{C}{\gamma^2}x^2+
n(n+1)\hbar^2\frac{1}{x^2}$
with $n=1$ or $n=2$ in these two cases, for which $A_-=\hbar\left(\frac{d}{dx}+\frac{n}{x}\right)$
and $A_+=\hbar\left(\frac{d}{dx}+\frac{n-1}{x}\right)$.
So, for $C<0$, $C>0$ and $C=0$, 
$V_-$ corresponds to the harmonic oscillator potential,
the inverted oscillator potential, or to the free particle zero potential  
which are corrected by 
the additive inverse square term.
In the case $C<0$ the system $\hat{H}_-$ corresponds to
the isotonic oscillator.
In the last case $C=0$, the partner $V_-$ corresponds to 
the potential of the two-body Calogero system
 with special values of the coupling constant \cite{COP}.
A dual supersymmetric system with the first order supercharges 
is given by the superpotential 
$\mathcal{W}=\frac{\gamma}{2z}-\kappa$, 
and the partner potentials 
$U_\pm(z)=\frac{\gamma}{2}(\frac{\gamma}{2}-1)\frac{1}{z^2}
-\gamma\kappa\frac{1}{z}+\kappa^2$. 
If  one considers $x$  as a radial variable, then 
the picture can be understood in the context of the 
well known relationship between the Coulomb and oscillator 
problems \cite{CKSb,Kostelecky}.

\vskip0.1cm

$\bullet$ Take now $W=\gamma\tanh \kappa x$.  Then 
$w_\pm=(\gamma\mp\frac{1}{2}\hbar\kappa)\tanh\kappa x\pm
\frac{1}{2}\hbar\kappa\coth \kappa x$ and 
$\Delta=-\hbar^2\kappa^2\frac{1}{\cosh^2\kappa x}
\left(1+\frac{1}{4\sinh^2\kappa x}\right)<0$.
With the choice $C=-\frac{1}{4}\hbar^2\gamma^2\kappa^2$,
both potentials $V_\pm(x)$ will be regular 
functions on the real line,
$V_\mp=\gamma^2+\frac{1}{4}\hbar^2\kappa^2
-\frac{\beta_\mp}{\cosh^2\kappa x}$, where 
$\beta_\mp=\left((\gamma\pm \hbar\kappa)^2-
\frac{1}{4}\hbar^2\kappa^2\right)$.
When one of the coefficients $\beta_-$ or $\beta_+$ 
is equal to zero,  one of the corresponding 
superpartner systems will be the free  particle.
For $\gamma=-\frac{1}{2}\hbar\kappa$, we have 
$V_-=\frac{1}{2}\hbar^2\kappa^2$ and 
$V_+=\frac{1}{2}\hbar^2\kappa^2 -
2\hbar^2
\kappa^2\frac{1}{\cosh^2\kappa x}$.
This $V_+$ corresponds to the reflectionless P\"oschl-Teller
system with one bound state in the spectrum \cite{CJPAdS}.
Both first order operators $A_-$ and $A_+$ 
in this case are singular,
$A_-=\hbar\left(\frac{d}{dx}-\frac{1}{2}\kappa\coth\kappa x\right)$,
$A_+=\hbar\left(\frac{d}{dx}+\frac{1}{2}\kappa\coth\kappa x
-\kappa\tanh\kappa x\right)$.
In spite of this, the second order intertwining operator
(\ref{q+AA}) is regular and 
can be presented in the factorized form
\be\label{q+regular}
	2\hat{q}_+=\hbar^2\left(
	\frac{d}{dx}-\kappa\tanh\kappa x\right)\frac{d}{dx}\,.
\ee
Let us find the operators $\mathcal{A}_\pm$ and their conjugate operators
which factorize the superpartner systems.   
In correspondence with the relation $\alpha^2=-16C$, see Eq. (\ref{abalp}),
we have two possibilities: $\alpha=\hbar^2\kappa^2$ and $\alpha=-\hbar^2\kappa^2$.
The first choice gives $\lambda(x)=\frac{1}{2}\hbar\kappa\coth\kappa x$,
and we obtain the factorizing operators in the form 
$$
\mathcal{A}_-=\hbar\left(\frac{d}{dx}-\kappa\coth\kappa x\right)\,,\qquad
\mathcal{A}_+=\hbar\left(\frac{d}{dx}+\kappa\coth\kappa x-\kappa\tanh\kappa x\right)\,.
$$
Both these operators similarly to  $A_-$ and $A_+$ are singular at $x=0$.
In correspondence with relations (\ref{intert1}) we have 
$\mathcal{A}_-^\dagger \mathcal{A}_-=\hbar^2(-\frac{d^2}{dx^2}+\kappa^2)$,
$\mathcal{A}_+ \mathcal{A}_+^\dagger=\hbar^2(-\frac{d^2}{dx^2}-2\kappa^2\frac{1}{\cosh^2\kappa x})$,
and the product $\mathcal{A}_+\mathcal{A}_-$ gives us the intertwining 
second order operator which can be presented in the equivalent form (\ref{q+regular}). 
The intermediate system here
is  singular, $\mathcal{A}_-\mathcal{A}_-^\dagger=\hbar^2(-\frac{d^2}{dx^2}
+\kappa^2+2\kappa^2\frac{1}{\sinh^2\kappa x})$,
see Eq. (\ref{intert3}).  

For $\alpha=-\hbar^2\kappa^2$, we have 
$\lambda(x)=-\frac{1}{2}\hbar\kappa\coth\kappa x$,
that gives us, instead, the non-singular factorizing operators 
$$
\mathcal{A}_-=\hbar\frac{d}{dx}\,,\qquad
\mathcal{A}_+=\hbar\left(\frac{d}{dx}-\kappa\tanh\kappa x\right)\,.
$$
It is these operators that factorize 
 the second order intertwining operator in  (\ref{q+regular}).
The intermediate system in this case is the same free particle
with the Hamiltonian operator shifted for another additive constant,
$\mathcal{A}_-\mathcal{A}_-^\dagger=-\hbar^2\frac{d^2}{dx^2}
=\mathcal{A}^\dagger_+\mathcal{A}_+-
\hbar^2\kappa^2$. 
This reflects also the fact that the pair of the systems 
$\hat{H}_-$ and $\hat{H}_+$  being reflectionless, 
in the present  case is described simultaneously  by supersymmetry 
with the linear supercharges constructed from the  
first order intertwining operators\footnote{Being reflectionless systems,
they are characterized also by the integrals of motion
which are $\hat{p}=-\hbar\frac{d}{dx}$ and 
 $\mathcal{A}_-\hat{p}\mathcal{A}_-^\dagger$~\cite{CJPAdS}.}
 $\mathcal{A}_+$ and $\mathcal{A}_+^\dagger$.
In correspondence with this property,
in the construction of the associated dual system with the 
first order supercharges the additive constant in superpotential
can be chosen in such a way that
we obtain $\mathcal{W}=-\hbar\kappa\tanh z$, see Section \ref{Metamor},
and the associated supersymmetric system will coincide exactly
with our system (with the variable $x$ changed for $z$).

Another choice of the parameter $\gamma=\frac{3}{2}\hbar\kappa$,
which gives the  free particle potential $V_+=\frac{5}{2}\hbar^2\kappa^2$ and
$V_-=\frac{5}{2}\hbar^2\kappa^2-6\hbar^2\kappa^2\frac{1}{\cosh^2\kappa x}$,
is also interesting.  In this case the $\hat{H}_-$ is the reflectionless P\"oschl-Teller 
system with two bound states in the spectrum \cite{CJPAdS},  
and
both operators $A_-$ and $A_+$, again,  are singular,
$A_-=\hbar\left(\frac{d}{dx}+2\kappa\tanh\kappa x-\frac{1}{2}\kappa\coth\kappa x\right)$,
$A_+=\hbar\left(\frac{d}{dx}+\kappa\tanh\kappa x+\frac{1}{2}\kappa\coth\kappa x\right)$.
Here the choice  $\alpha=3\hbar^2\kappa^2$ leads to  nonsingular
factorizing operators,
\be\label{AAnomer1}
	\mathcal{A}_-=\hbar\left(\frac{d}{dx}+2\kappa\tanh\kappa x\right)\,,\qquad
	\mathcal{A}_+=\hbar\left(\frac{d}{dx}+\kappa\tanh\kappa x\right)\,,
\ee
in terms of which we have 
$\mathcal{A}_+\mathcal{A}_+^\dagger=-\hbar^2\frac{d^2}{dx^2}+\hbar^2\kappa^2$,
$\mathcal{A}_-^\dagger\mathcal{A}_-=-\hbar^2\frac{d^2}{dx^2}+4\hbar^2\kappa^2
-6\hbar^2\kappa^2\frac{1}{\cosh^2\kappa x}$.
The Hamiltonians are intertwined by the second order
operator $2\hat{q}_+=\mathcal{A}_+\mathcal{A}_-$.
The intermediate system in this case is the reflectionless 
P\"oschl-Teller system with one bound state,
$\mathcal{A}_+^\dagger\mathcal{A}_+=-\hbar^2\frac{d^2}{dx^2}+\hbar^2\kappa^2
-2\hbar^2\kappa^2\frac{1}{\cosh^2\kappa x}$.

Another choice  $\alpha=-3\hbar^2\kappa^2$ leads to
the singular
factorizing operators,
\be\label{AAnomer2}
	\mathcal{A}_-=\hbar\left(\frac{d}{dx}+2\kappa\tanh\kappa x-\kappa\coth\kappa x\right)\,,\qquad
	\mathcal{A}_+=\hbar\left(\frac{d}{dx}+2\kappa\coth2\kappa x\right)\,.
\ee
Here  
$\mathcal{A}_+\mathcal{A}_+^\dagger=-\hbar^2\frac{d^2}{dx^2}+4\hbar^2\kappa^2$,
$\mathcal{A}_-^\dagger\mathcal{A}_-=-\hbar^2\frac{d^2}{dx^2}+\hbar^2\kappa^2
-6\hbar^2\kappa^2\frac{1}{\cosh^2\kappa x}$.
The peculiarity  is that for intermediate, virtual system 
we have here
$\mathcal{A}_+^\dagger\mathcal{A}_+=4\hbar^2\left(
\frac{d^2}{d\xi^2}+\hbar^2\kappa^2+2\kappa^2\frac{1}{\sinh^2\kappa\xi}\right)$,
where $\xi=2x$. The intermediate system
is also finite-gap\footnote{In the sense that for it,
a nontrivial Lax-Novikov integral exists and is given 
by $\mathcal{A}_-^\dagger \hat{p}\mathcal{A}_+$, but this integral and  the 
Hamiltonian $\hat{H}_\otimes$ are singular operators.},
but to see this explicitly requires rescaling of the variable. 
Since $\hat{H}_-$ and 
$\hat{H}_+$  are reflectionless, 
the system composed from them 
 is characterized also by the supercharges which are
differential operators of the third order constructed from 
the intertwining operator $\hbar\hat{q}_+\frac{d}{dx}$ and its conjugate,
where $2\hat{q}_+=\mathcal{A}_+\mathcal{A}_-$
with $\mathcal{A}_\pm$ given either by (\ref{AAnomer1}) or 
 (\ref{AAnomer2}). 
The reflectionless P\"oschl-Teller subsystem $\hat{H}_-$ is characterized 
also by a nontrivial Lax-Novikov integral which is
the fifth order differential operator 
$\mathcal{A}_-^\dagger\mathcal{A}_+^\dagger\hat{p}
\mathcal{A}_+\mathcal{A}_-$ \cite{CJPAdS}.

\section{Conclusion}\label{Conclusion}
We have investigated the problem of the quantum anomaly
in the systems with the second order supersymmetry and found 
that it is treated universally by modifying the quantization
prescription for the quadratic in momentum terms
both in the  Hamiltonian and in the supercharges. 
Such a modification generates a specific 
quadratic in Planck constant term which in a hidden 
form has a nature of the Schwarzian derivative.
Though the quantization prescription we employed
is somewhat similar to that which is used  in the quantum problem 
of a particle in a curved space \cite{deWitt,deAlf},
there is an essential difference.
Namely,
in the case of  the particle  in  curved space described 
by external metric $g_{\mu\nu}(x)$, 
the classical kinetic term has a form 
$p_\mu g^{\mu\nu}p_\nu$, and 
there appears a problem  
of ordering ambiguity under construction of its  quantum analog
due to non-commutativity of the momentum operators
with metric tensor. The ordering problem is fixed there by requiring 
the preservation of  the invariance of the theory under general 
coordinate transformations, that is achieved by 
taking the quantization prescription for the kinetic term
in the form $p^-_\mu g^{\mu\nu}p^+_\nu$,
where $p^\pm_\mu=g^{\pm 1/4}p_\mu g^{\mp 1/4}$ \cite{deWitt}.
In Hamiltonian of the classical system with the second order supercharges 
that we considered, the kinetic term appears in the form
$p^2$, and so, there is no ordering ambiguity 
in it under transition to the quantum case. However,  if we take
its quantum analog in the form $-\hbar^2\frac{d^2}{dx^2}$,
with such a natural quantization prescription the second order
supersymmetry will be lost at the quantum level and we 
face the problem of the quantum anomaly.
We `cure'  this problem by introducing at the classical level
a  fictitious similarity transformation (\ref{p2equiv}) 
into the kinetic term of the Hamiltonian and 
a similar fictitious  transformation is introduced 
into the classical supercharges. Fixing then 
the generating function of  the  similarity  transformation 
in a special 
form in terms of the superpotential and taking 
the ordering prescription similarly to the case of the particle in  the curved
 space, we recuperate  the second order supersymmetry at the
 quantum level by generating 
 the necessary term (\ref{Delta(W)}), which
 is quadratic in the quantum constant $\hbar$
 and can be presented in the form
 of the Schwarzian derivative 
 (\ref{DeltaSchwarz}).
It would be interesting to analyze  the indicated  similarity 
with the particle in the curved space in more detail. This could be helpful
in investigation of  the problem of the quantum anomaly 
in the systems with nonlinear supersymmetries of the order
higher than two, particularly, of the order three \cite{IofNis}.
This analogy as well as a noted at the beginning similarity
with finite W-algebras may indicate that the 
one-dimensional quantum mechanical
systems with higher order supersymmetry 
could be generated by application of
some  reduction procedure 
to a particle with linear supersymmetry 
but `living' in a space of dimension $d>1$.
In such a case the peculiar and seemingly unnatural 
quantization prescription
used  here to maintain  the second order
supersymmetry could find a reasonable 
explanation from the point of view of  quantization in 
a higher-dimensional
curved space.

When the constant $C$ that appears in the structure of
the Hamiltonian operators  (\ref{H+-AA}) and supercharges 
(\ref{q+AA})  is positive, 
the partner Hamiltonians are factorized in terms of the  first order operators 
which are not Hermitian conjugate, and in this case 
the factorization relations (\ref{intert1}) involve a complex shift parameter.
It would be interesting to investigate  the properties of such 
a class of supersymmetric systems in more detail,
particularly, from the perspective  
of their possible physical applications.  

Finally we note that the phenomenon 
of the coupling-constant metarmorphosis and related 
duality between the systems with the second and the first order 
supersymmetries that we observed in Section \ref{Metamor}
deserves a further investigation.
This is interesting particularly from the point of view 
of the finite-gap and reflectionless 
systems which reveal a peculiar, exotic
supersymmetric structure associated with the presence in 
them of Lax-Novikov integrals 
\cite{CJNP,CJPAdS,CorNiePly,CorrDunPl,AGP1,AnMP2,ACJGP}.
The results presented here may also be useful
for better understanding of a peculiar nature 
of the  recently discovered class 
of the quantum mechanical exactly solvable 
systems related to exceptional orthogonal polynomials 
\cite{Odake,Quesne11,DavGom,Car,FelSmi}.

\vskip0.2cm
\noindent \textbf{Acknowledgements.}
The work has been partially supported by FONDECYT Grant
No. 1130017 and  Proyecto  Basal USA1555.

\section{Apendix. Factorizing operators}\label{Appendix}

Here we show that the first order operators
(\ref{calA}) corresponding to factorization  (\ref{intert1})
are not fixed completely  by relations  (\ref{abalp}) 
and by the choice of the sign of parameter $\alpha$,
and we find their one-parameter generalizations which are doing 
the same job.
This is analogous to a simple picture that we have 
in the case of the free particle whose Hamiltonian
$\hat{H}_0=-\frac{d^2}{dx^2}$  admits a family of
factorizing operators, $\hat{H}_0=\mathcal{A}^\sharp_\zeta\mathcal{A}_\zeta$,
$\mathcal{A}_\zeta=\frac{d}{dx}+\frac{1}{x+\zeta}$,
$\mathcal{A}_\zeta^\sharp=-\frac{d}{dx}+\frac{1}{x+\zeta}$,
where  $\zeta$ is an arbitrary complex parameter (of 
a dimension of length), and  $\mathcal{A}_\infty=
-\mathcal{A}_\infty^\sharp=\frac{d}{dx}$.
Below, as in  example of the free particle,
we are not preoccupied with  the questions of 
regularity of operators and their conjugation 
properties.

Let $\psi(x)$ be a  solution of the stationary Schr\"odinger equation
$(\hat{H}-E)\psi=0$, where $E$ in general  
can be complex.   
We have then a natural factorization
$(\hat{H}-E)=\mathcal{A}^\sharp\mathcal{A}$
with 
\be\label{AAdef1}
	\mathcal{A}=\psi(x)\frac{d}{dx}\frac{1}{\psi(x)}=
	\frac{d}{dx}-{w}(x)\,,\quad 
	\mathcal{A}^\sharp=-
	\frac{1}{\psi(x)}\frac{d}{dx}\psi(x)=
	-\frac{d}{dx}-{w}(x)\,,
\ee
and 
${w}(x)=(\ln \psi(x))'$.
A  linear independent solution of the second order
differential equation $(\hat{H}-E)\psi=0$ can be taken in the form 
$\psi(x;C)=\psi(x)\left(1+C\int_{x_0}^x\frac{ds}{\psi^2(s)}\right)$,
where $x_0\in \R$ is some fixed point and $C\neq0$ is an
arbitrary complex constant. The action of $\mathcal{A}$ on
$\psi(x;C)$ produces zero mode of $\mathcal{A}^\sharp$,
$\mathcal{A}\psi(x;C)=\frac{C}{\psi(x)}$,
and as a result,  $(\hat{H}-E)\psi(x;C)=
\mathcal{A}^\sharp\mathcal{A}\psi(x;C)=0$.
Obviously, instead of $\mathcal{A}$ and $\mathcal{A}^\sharp$, 
we can define the first order 
factorization operators $\mathcal{A}(C)$ and $\mathcal{A}^\sharp(C)$
given by relations (\ref{AAdef1}) but with $\psi(x)$ changed for $\psi(x;C)$.
As a result, we obtain  $(\hat{H}-E)=\mathcal{A}^\sharp(C)\mathcal{A}(C)$
with 
\be\label{AACdef}
	\mathcal{A}(C)=\frac{d}{dx}-w(x;C)\,,\qquad
	\mathcal{A}^\sharp(C)=-\frac{d}{dx}-w(x;C)\,,
\ee
where
\be\label{WdefC}
	{w}(x;C)=w(x)+\frac{C\exp(-2\Xi(x))}{1+
	C\int_{x_0}^x\exp(-2\Xi(s))ds}\,,
\ee
and $\Xi(x)=\int_{x_0}^x{w}(s)ds$
so that $\exp(-2\Xi(x))=\frac{1}{\psi^2(x)}$. 
At $C=0$, ${w}(x;C)$ transforms into ${w}(x)$,
and (\ref{AACdef}) transforms into (\ref{AAdef1}).
The partner Sch\"odinger operator $\mathcal{A}(C)\mathcal{A}^\sharp(C)$
here is $C$-dependent. For instance,  in the 
above mentioned example of the free particle we have
$\mathcal{A}_\zeta\mathcal{A}^\sharp_\zeta=-\frac{d^2}{dx^2}
+\frac{2}{(x+\zeta)^2}$.

Using Eqs. (\ref{calA}) and (\ref{intert1}),
one can find zero modes $\Psi_-(x)$ and $\Psi_+(x)$
 of operators $\mathcal{A}_-$ and
$\mathcal{A}_+^\sharp$  being also zero modes
of $2(\hat{H}_--b)$ and $2(\hat{H}_+-b)-\alpha$,
respectively.  Then, following the same reasoning, 
as a generalization of 
$\mathcal{A}_-$ and  $\mathcal{A}_-^\sharp$ we obtain
 the factorizing operators $\mathcal{A}_-(C_-)=\mathcal{A}_-+
 \Lambda_-(W;C_-)$ and 
 $\mathcal{A}_-^\sharp(C_-)=\mathcal{A}_-^\sharp+
 \Lambda_-(W;C_-)$, 
 where $C_-\in \C$, and
 \be\label{Lam-Om}
 	 \Lambda_-(W;C_-)=\frac{C_-\Omega_-(x)}{
 	 1+C_-\int_{x_0}^x\Omega_-(s)ds}\,,  \qquad
 	\Omega_-(x)=\frac{1}{W(x)}\exp\left(2\int_{x_0}^x(W(s)-\lambda(s))ds\right).
\ee
 For these operators we have
 $2(\hat{H}_--b)=\mathcal{A}_-^\sharp(C_-)\mathcal{A}_-(C_-)$.
 
As a generalization of 
$\mathcal{A}_+$ and  $\mathcal{A}_+^\sharp$ we obtain
  $\mathcal{A}_+(C_+)=\mathcal{A}_++
 \Lambda_+(W;C_+)$ and 
 $\mathcal{A}_+^\sharp(C_+)=\mathcal{A}_+^\sharp+
 \Lambda_+(W;C_+)$, 
 with $C_+\in\C$,  and 
 \be\label{Lam+Om}
	  \Lambda_+(W;C_+)=\frac{C_+\Omega_+(x)}{
	  1+C_+\int_{x_0}^x\Omega_+(s)ds}\,,  \qquad
 	\Omega_+(x)=\frac{1}{W(x)}\exp\left(-2\int_{x_0}^x(W(s)+\lambda(s))ds\right).
\ee
We have
 $2(\hat{H}_+-b)-\alpha=
 \mathcal{A}_+(C_+)\mathcal{A}_+^\sharp(C_+)$.
 Functions $\Omega_-(x)$ and  $\Omega_+(x)$ 
 that appear in (\ref{Lam-Om}) and (\ref{Lam+Om})
 are nothing else as the inverse squares of zero modes of the first order
 operators $ \mathcal{A}_-$ and $ \mathcal{A}_+^\sharp$,
 $\Omega_\mp(x)=\frac{1}{\Psi^2_\mp(x)}$.



\end{document}